\documentclass[a4paper,fleqn,usenatbib]{mnras}

\usepackage{newtxtext,newtxmath}
\usepackage{adjustbox}
\usepackage{hyperref}
\usepackage{ragged2e}  

\usepackage[T1]{fontenc}
\usepackage{ae,aecompl}

\usepackage{graphicx}	
\usepackage{amsmath}	

\title[Study of central intensity ratio]{Study of central intensity ratio of early-type galaxies from low density environment}

\author[Sruthi $\&$ Ravikumar]{
K. Sruthi,$^{1}$\thanks{E-mail: sruthiyatheendradas@gmail.com}
C. D. Ravikumar$^{1}$
\\
$^{1}$Department of Physics, University of Calicut, Malappuram-673635, India.\\
}

\date{Accepted XXX. Received YYY; in original form ZZZ}

\pubyear{2020}

\begin{document}
\label{firstpage}
\pagerange{\pageref{firstpage}--\pageref{lastpage}}
\maketitle

\begin{abstract}
We present correlations involving central intensity ratio (CIR) of 52 early type galaxies, including 24 ellipticals and 28 lenticulars, selected from low density environment in the nearby (< 30 Mpc) universe.  CIR is found to be negatively and significantly correlated with the mass of the central super massive black hole, central velocity dispersion, absolute B band magnitude, stellar bulge mass and central Mg$_{2}$ index of the host galaxy. The study proposes the use of CIR as a simple, fast and efficient photometric tool for exploring the co-evolution scenario existing in galaxies.

\end{abstract}




\section{Introduction}
Early-type galaxies (hereafter ETGs), a category encompassing lenticular and elliptical galaxies are having uniform nature in some of their properties and described as red, passive/retired and morphologically featureless systems at the end of galaxy evolution. They possess smooth light profiles, very low gas and dust content. The spectroscopic and photometric observations of nearby ETGs reveal that, they obey tight scaling relations such as color-magnitude relation \citep{1978ApJ...223..707S, 1980ApJ...237..692L, 1992MNRAS.254..601B} , Mg$_2$ - $\sigma$ relation \citep{1981MNRAS.196..381T, 1993ApJ...411..153B, 1999MNRAS.303..813C}, Fundamental plane relation  \citep{1987ApJ...313...59D,1987ApJ...313...42D} which are interpreted as an evidence for a very homogeneous population of early-type galaxies. With increasing galaxy mass, ETGs are older, more metal rich and exhibit higher fractions of $\alpha$ to Fe peak elements \citep[e.g.][]{2005ApJ...621..673T,2012MNRAS.421.1908J}. Environment plays a substantial role in driving galaxy evolution, although the details of the physical processes involved are still poorly understood. In the nearby universe, galaxy properties such as star formation rate (SFR), morphology, stellar mass and nuclear activity are strongly correlated with the surrounding galaxy density \citep[e.g.][]{1974ApJ...194....1O, 1980ApJ...236..351D, 1998ApJ...499..589H, 2004MNRAS.353..713K, 2006MNRAS.373..469B, 2019ApJ...874..140A}.

Massive ETGs are believed to host a supermassive black hole (SMBH) at their  centres \citep{1995ARA&A..33..581K}. As samples with robust measurements of mass of SMBH have increased, several black hole - host galaxy correlations have been reported, including central velocity dispersion \citep{2000ApJ...539L...9F,2000ApJ...539L..13G,2002ApJ...574..740T,2013ARA&A..51..511K,2016ApJ...818...47S,2019MNRAS.487.5764T,2019GReGr..51...65Z}, bulge mass/luminosity \citep{1995ARA&A..33..581K,1998AJ....115.2285M, 2002MNRAS.331..795M, 2003ApJ...589L..21M, 2004ApJ...604L..89H, 2013ApJ...767...13S, 2019ApJ...887..245S}, light concentration \citep{2001ApJ...563L..11G, 2007ApJ...655...77G}, bulge kinetic energy \citep{2012A&A...537A..48M, 2009ApJ...703.1502F}, the dark matter halo \citep{2002ApJ...578...90F}, the S\'ersic index \citep{2013MNRAS.434..387S}, the pitch angle \citep{2017MNRAS.471.2187D}, core radius \citep{2016ApJ...818...47S}, core size of massive ellipticals \citep{2016Natur.532..340T} and with the central intensity ratio \citep{2018MNRAS.477.2399A}. These scaling relationships are indicative of an intimate connection between SMBH growth and galaxy formation/evolution \citep{2000ApJ...539L...9F, 2008ApJ...686..815Y, 2013ARA&A..51..511K, 2016ApJ...818...47S, 2019MNRAS.488L.134L}. Although most massive galaxies in the local universe host SMBH in their centres, active periods of accretion on to such objects, seen as active galactic nuclei (AGNs),  are observed in less than a few percent of them \citep{2019MNRAS.487.5764T}.    Models on galaxy formation have shown that AGN feedback plays an important role in shaping many properties of massive galaxies \citep[e.g.][]{2015MNRAS.449.4105C, 2017MNRAS.465.3291W} and it is widely considered as the most plausible mechanism for quenching in massive galaxies \citep{2018NatAs...2..695M}.

Following \citet{1993MNRAS.264..832D}, the idea of central concentration of light was first presented by \citet{1994ApJ...432...75A} with an interest to trace stellar population. \citet{2001MNRAS.326..869T} proposed a galaxy concentration index practically independent of the image exposure depth related to the S\'{e}rsic index \textit{n} \citep{1968adga.book.....S}. \citet{2001ApJ...563L..11G} defined concentration index  as the ratio of flux inside a fixed fraction of half-light radius to that within the half-light radius and reported a correlation between light concentration and black hole mass. Concentration of light in the galaxies was also used to trace past evolutionary history of galaxy formation \citep{2003ApJS..147....1C}. \citet{2018MNRAS.477.2399A} defined central intensity ratio (CIR, see Section \ref{sec:central}) to measure the variation of light intensity at the very centre of the galaxy image. The simple definition of CIR helps avoid any dependence on a form following central intensity, \textit{I}(0) (i.e. surface brightness at a radial distance r, \textit{I}(r) = \textit{I}(0) \textit{f}(r) where \textit{f}(r) is a function of r). Also, the definition boosts any addition to (or subtraction from) the central intensity \textit{I}(0). Further, simple Monte Carlo simulations show remarkable stability against variations in distance and orientation of ellipsoidal systems in the nearby universe \citep{2018MNRAS.477.2399A,2020RAA....20...15A}.  Pseudo-bulge hosting galaxies were found to be outliers  in majority of the correlations involving CIR  \citep{2018MNRAS.477.2399A}. In this paper we study the correlations of CIR for a carefully selected sample of nearby field ETGs.

The paper is organized as follows. Section \ref{sec:data} describes the sample selection, observations, data reduction and calculation of central intensity ratio. Section \ref{sec:correlations} presents correlations involving CIR with different host galaxy properties.  We present discussion of results and conclusion in Section \ref{sec:discussion}.
\section{THE DATA}
\label{sec:data}
Our sample consists of a representative sample of nearby (<30 Mpc) field ETGs.  It consists of all field ETGs from \citet{2016ApJ...829...20W} with \textit{HST} observation. A galaxy is identified as a field galaxy if there is no sizeable neighbours within 1.5 magnitudes, 0.26 Mpc in the plane of the sky and 500 km s$^{-1}$ in recession velocity. It is highly desirable to have all observations in one filter in order to have homogeneity in the data and we looked for observations with \textit{WFPC2} camera  in \textit{F814} filter. However, we noticed that for the different \textit{HST} filters available for a galaxy, the difference in CIR estimations were too small to affect the correlations discussed in this paper.  Hence we used all available observations of \textit{HST} using cameras \textit{WFPC2} and \textit{ACS} from filters \textit{F547} to \textit{F850} for want of reasonable sample size. Whenever observations in the filter \textit{F814} (using \textit{WFPC2}) were absent, images from the nearest available filter was used. When no \textit{WFPC2} observation was available, we used  \textit{ACS} images from the nearest filter.
We excluded galaxy images with bad pixels in their central 3 arcsec region. Also, we avoided the elliptical galaxy NGC 3125 owing to its disturbed morphology. The sample thus created consists of 52 ETGs including 24 ellipticals and 28 lenticulars. The sample is listed in Table \ref{tab:table1}. 

\subsection{Central Intensity Ratio}
\label{sec:central}
 \citet{2018MNRAS.477.2399A} defined CIR as,
\begin{equation}
CIR = \frac{I_{1}}{I_{2} - I_{1}} = \frac{10^{0.4(m_{2}-m_{1})}}{1-10^{0.4(m_{2}-m_{1})}}
\end{equation}
where I$_{1}$ and I$_{2}$ are the intensities and m$_{1}$ and m$_{2}$ are the magnitudes within the inner and outer apertures respectively. Aperture photometry (MAG$\_$APER) using Source extractor
\citep[SEXTRACTOR,][]{1996A&AS..117..393B} was carried out at the optical centre of the galaxy image taking the inner and outer apertures as 1.5  and 3 arcsec, respectively. The inner radius is selected to contain the effects of the point spread function (PSF) in the  \textit{HST} images and the outer radius is chosen to be smaller than the half light radii of the sample galaxies, so that the estimation of CIR is reasonably stable against differences in size and surface brightness profile \citep{2018MNRAS.477.2399A}.
\begin{figure*}
\includegraphics[width=1.0\textwidth]{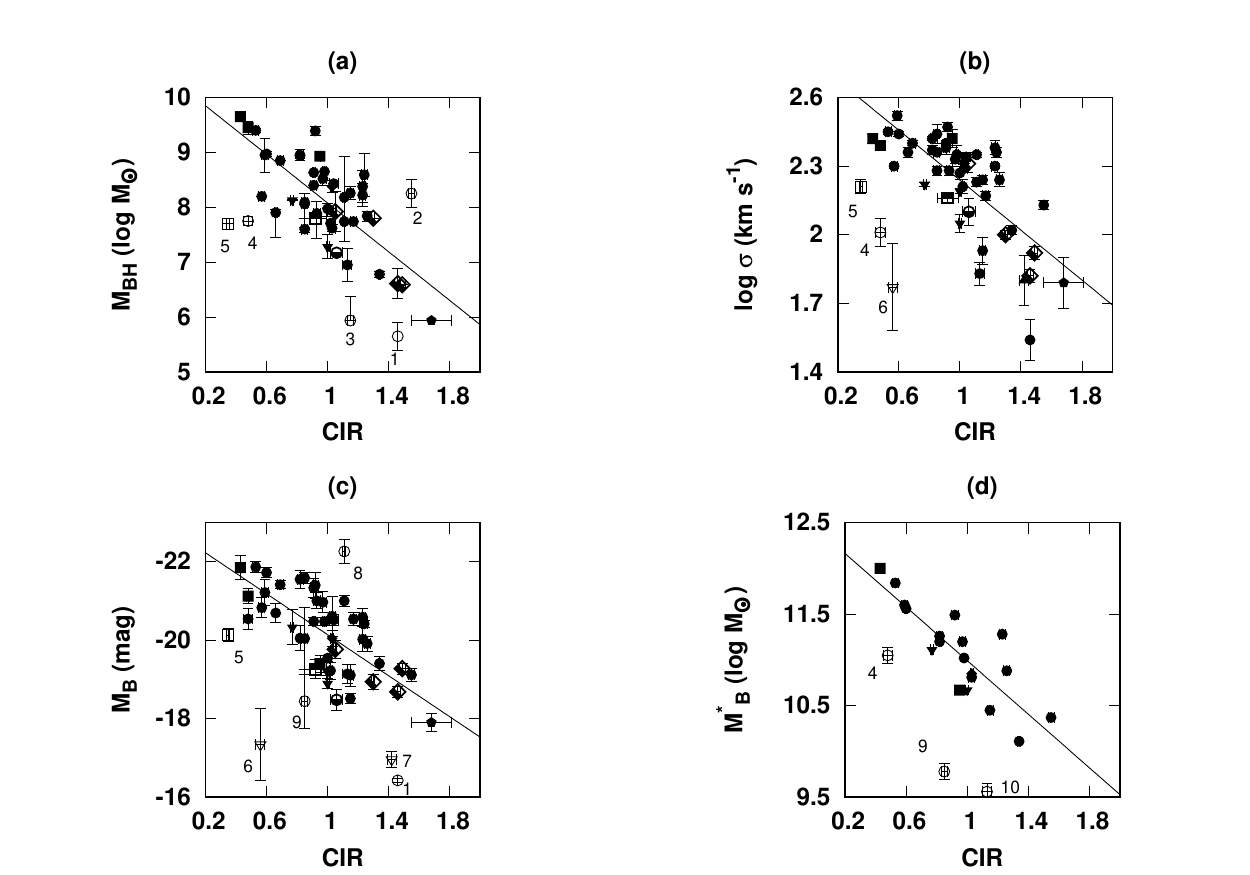}
\caption{Correlation between central intensity ratio and (a) mass of the SMBH of the host galaxy (M$_{BH}$), (b) central velocity dispersion ($\sigma$), (c) Absolute B band magnitude of the host galaxy  (M$_{B}$) and  (d) stellar bulge mass (M$^{*}_{B}$) of the galaxy. Each observation is represented by different symbols. Filled circles denote WFPC2-F814 observation, filled squares denote ACS-F814 observation, inverted filled triangle represent ACS-F850 observation, filled diamonds represent WFPC2-F702 observation, filled pentagons denote WFPC2-F656 observation, partially filled diamonds denote WFPC2-F606 observation, partially filled squares denote WFPC2-F658 observation, partially filled circles represent WFPC2-F547 observation. Open symbols corresponding to each observation is used to denote outlier galaxies. Number codes used to show outliers  are as follows: (1)  NGC 0404, (2)  NGC 3377, (3)  NGC 4150, (4)  NGC 5128, (5)  NGC 5866, (6) IC 3773, (7)  NGC 0855, (8) NGC 1316, (9)  NGC 2787, (10)  NGC 7457.}
\label{fig:figure1}
\end{figure*}

\begin{figure}
	\includegraphics[width=\columnwidth]{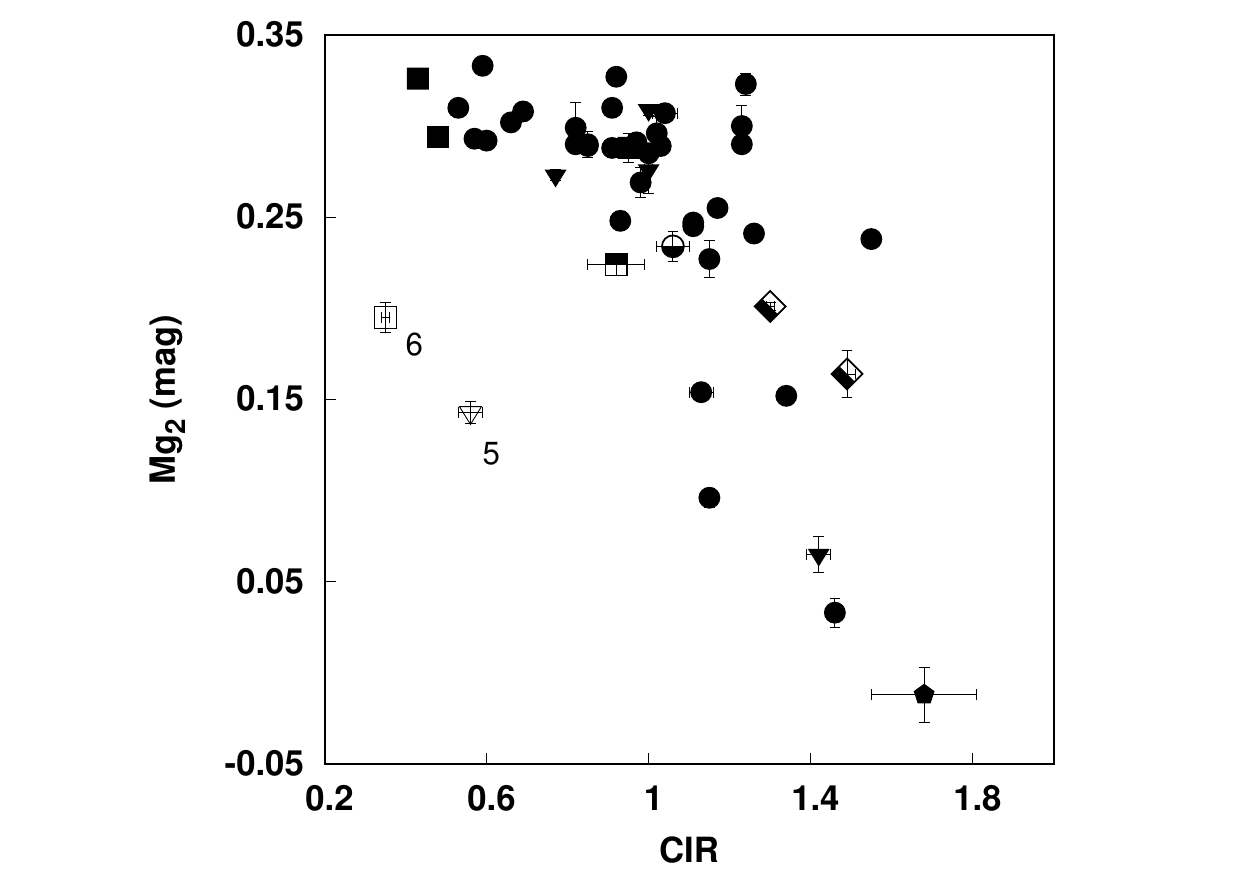}
    \caption{CIR versus central Mg$_{2}$ index of the host galaxy. The symbols are as shown in Figure \ref{fig:figure1}.}
    \label{fig:figure2}
\end{figure}

\begin{table*}
\caption{Table 1 lists the sample galaxy properties. Name of the galaxy (column 1), Distance of the galaxy taken from \citet{2016ApJ...829...20W} (2), Details of HST observation (3), Morphology of galaxy (E stands for elliptical and S0 stands for lenticular) taken from \citet{2016ApJ...829...20W} (4), CIR computed in the corresponding filter (5), uncertainty in the measurement of CIR (6), Mass of SMBH (7) and corresponding references (8), central velocity dispersion (9) central Mg$_{2}$ index (10) and absolute B band magnitude (11) of the galaxy adopted from HyperLEDA, stellar bulge mass (12) of the galaxy adopted from \citet{2013ApJ...764..184M} and \citet{2013ARA&A..51..511K} }
\begin{tabular}{lccccccccccc}
\hline
\multicolumn{1}{c}{GALAXY} & Distance & HST & Morphology & CIR & $\bigtriangleup_{CIR}$ & M$_{BH}$ & Ref. & $\sigma$ & Mg$_{2}$ & M$_{B}$ & M$^{*}_{B}$ \\ 
\multicolumn{1}{c}{} & (Mpc) & observation &  &  &  & (log M$_{\odot}$) &  & (km s$^{-1}$) & (mag) & (mag) & (log M$_{\odot}$) \\ 
\hline
IC1459 & 29.24 & WFPC2\_F814 & E & 0.92 & 0.02 & 9.39 & 1 & 296.11 & 0.327 & -21.403 & 11.49 \\ 
IC3773 & 14.5 & ACS\_F850 & E & 0.56 & 0.03 & - & - & 59.49 & 0.143 & -17.342 & - \\ 
NGC0404  & 3.27 & WFPC2\_F814 & S0 & 1.46 & 0.00 & 5.65 & 13 & 34.59 & 0.033 & -16.428 & - \\ 
NGC0524  & 23.99 & WFPC2\_F814 & S0 & 0.82 & 0.02 & 8.94 & 1 & 236.69 & 0.299 & -21.554 & 11.26 \\ 
NGC0821 & 24.1 & WFPC2\_F814 & E & 1.23 & 0.02 & 8.22 & 1 & 197.73 & 0.29 & -20.59 & 11.28 \\ 
NGC0855 & 9.73 & ACS\_F850 & E & 1.42 & 0.03 & - & - & 62.74 & 0.065 & -16.96 & - \\ 
NGC1023  & 11.43 & WFPC2\_F814 & S0 & 1.03 & 0.02 & 7.62 & 1 & 197.1 & 0.289 & -20.61 & 10.81 \\ 
NGC1316  & 21.48 & WFPC2\_F814 & S0 & 1.11 & 0.01 & 8.18 & 10 & 223.06 & 0.245 & -22.264 & - \\ 
NGC1399 & 19.95 & WFPC2\_F814 & E & 0.59 & 0.02 & 8.95 & 1 & 332.23 & 0.333 & -21.218 & 11.6 \\ 
NGC1404 & 20.99 & WFPC2\_F814 & E & 0.66 & 0.01 & 7.9 & 5 & 229.71 & 0.302 & -20.696 & - \\ 
NGC1407 & 28.84 & ACS\_F814 & E & 0.43 & 0.01 & 9.65 & 1 & 265.63 & 0.326 & -21.85 & 12 \\ 
NGC2787  & 7.48 & WFPC2\_F814 & S0 & 0.85 & 0.02 & 7.6 & 10 & 191.92 & 0.29 & -18.447 & 9.78\\ 
NGC3115  & 9.68 & WFPC2\_F814 & S0 & 0.82 & 0.01 & 8.95 & 1 & 260.23 & 0.29 & -20.052 & 11.2 \\
NGC3245 & 20.89 & WFPC2\_F702 & S0 & 1.03 & 0.01 & 8.38 & 1 & 207.03 & - & -20.049 & 10.85 \\ 
NGC3377 & 11.22 & WFPC2\_F814 & E & 1.55 & 0.02 & 8.25 & 1 & 136.1 & 0.238 & -19.12 & 10.37 \\ 
NGC3412 & 11.32 & WFPC2\_F606 & S0 & 1.3 & 0.01 & 7.8 & 15 & 99.46 & 0.201 & -18.941 & - \\ 
NGC3414  & 25.23 & WFPC2\_F814 & S0 & 1.23 & 0.02 & 8.38 & 11 & 237.55 & 0.3 & -20.028 & - \\ 
NGC3489  & 12.08 & WFPC2\_F814 & S0 & 1.34 & 0.01 & 6.78 & 1 & 104.18 & 0.152 & -19.408 & 10.11 \\ 
NGC3585 & 20.04 & WFPC2\_F814 & E & 0.97 & 0.02 & 8.52 & 1 & 214.2 & 0.291 & -20.969 & 11.2 \\ 
NGC3610 & 21.38 & WFPC2\_F814 & E & 1.11 & 0.01 & 7.74 & 4 & 168.27 & 0.247 & -21.003 & - \\ 
NGC3640 & 27.04 & WFPC2\_F814 & E & 0.93 & 0.02 & 7.89 & 8 & 191.43 & 0.248 & -21.005 & - \\ 
NGC3923 & 22.91 & ACS\_F814 & E & 0.48 & 0.01 & 9.45 & 1 & 245.55 & 0.294 & -21.123 & - \\ 
NGC3998 & 14.13 & ACS\_F814 & S0 & 0.95 & 0.00 & 8.93 & 1 & 265.09 & 0.288 & -19.404 & 10.67 \\ 
NGC4026  & 13.61 & WFPC2\_F814 & S0 & 1.15 & 0.02 & 8.26 & 1 & 173.49 & 0.227 & -19.11 & 10.45 \\ 
NGC4111 & 15 & WFPC2\_F658 & S0 & 0.92 & 0.07 & 7.8 & 3 & 146.2 & 0.224 & -19.263 & - \\ 
NGC4125 & 23.88 & WFPC2\_F814 & E & 0.91 & 0.02 & 8.4 & 3 & 239.78 & 0.288 & -21.337 & - \\ 
NGC4138 & 13.8 & WFPC2-F547 & S0 & 1.06 & 0.04 & 7.17 & 18 & 126.25 & 0.234 & -18.481 & - \\ 
NGC4143 & 15.92 & WFPC2\_F606 & S0 & 1.05 & 0.01 & 7.92 & 9 & 202.42 & - & -19.771 & - \\ 
NGC4150  & 13.74 & WFPC2\_F814 & S0 & 1.15 & 0.02 & 5.94 & 9 & 85 & 0.096 & -18.512 & - \\ 
NGC4203  & 15.14 & WFPC2\_F814 & S0 & 1.02 & 0.01 & 7.71 & 14 & 161.66 & 0.296 & -19.226 & - \\ 
NGC4365  & 20.42 & WFPC2\_F814 & E & 0.69 & 0.02 & 8.85 & 7 & 250.33 & 0.308 & -21.417 & - \\ 
NGC4374 & 18.37 & WFPC2\_F814 & E & 0.6 & 0.01 & 8.97 & 1 & 277.59 & 0.292 & -21.726 & 11.56 \\ 
NGC4406  & 17.14 & WFPC2\_F814 & E & 0.85 & 0.02 & 8.06 & 7 & 231.37 & 0.289 & -21.589 & - \\ 
NGC4459  & 16.14 & WFPC2\_F814 & S0 & 1.26 & 0.02 & 7.84 & 1 & 171.84 & 0.241 & -19.915 & 10.88 \\ 
NGC4472 & 16.29 & WFPC2\_F814 & E & 0.53 & 0.02 & 9.4 & 1 & 281.98 & 0.31 & -21.869 & 11.84 \\ 
NGC4494 & 17.06 & WFPC2\_F814 & E & 1.17 & 0.02 & 7.74 & 2 & 148.32 & 0.255 & -20.54 & - \\ 
NGC4526  & 16.9 & WFPC2\_F814 & S0 & 0.98 & 0.01 & 8.65 & 11 & 224.64 & 0.269 & -20.48 & 11.02 \\ 
NGC4552  & 15.35 & WFPC2\_F814 & E & 0.91 & 0.01 & 8.63 & 2 & 250.31 & 0.31 & -20.481 & - \\ 
NGC4564 & 15 & ACS\_F850 & E & 1 & 0.01 & 7.95 & 1 & 156.16 & 0.309 & -19.186 & 10.67 \\ 
NGC4570  & 16.8 & WFPC2\_F814 & S0 & 1 & 0.01 & 7.97 & 7 & 186.52 & 0.285 & -19.543 & - \\ 
NGC4578 & 18.54 & ACS\_F850 & S0 & 1 & 0.01 & 7.28 & 1 & 111.92 & 0.276 & -18.9 & - \\ 
NGC4589 & 21.98 & WFPC2\_F814 & E & 1.04 & 0.03 & 8.43 & 2 & 219.36 & 0.307 & -20.517 & - \\ 
NGC4621  & 18.28 & WFPC2\_F814 & E & 1.24 & 0.02 & 8.59 & 6 & 227.79 & 0.323 & -20.422 & - \\ 
NGC4636 & 14.66 & WFPC2\_F814 & E & 0.57 & 0.02 & 8.2 & 3 & 199.47 & 0.293 & -20.834 & - \\ 
NGC4697 & 11.75 & ACS\_F850 & E & 0.77 & 0.01 & 8.13 & 1 & 165.22 & 0.273 & -20.334 & 11.11 \\ 
NGC5102 & 4 & WFPC2\_F656 & S0 & 1.68 & 0.13 & 5.94 & 16 & 61.1 & -0.012 & -17.906 & - \\ 
NGC5128  & 4.21 & WFPC2\_F814 & S0 & 0.48 & 0.03 & 7.75 & 1 & 103.39 & - & -20.546 & 11.05 \\ 
NGC5273 & 16.52 & WFPC2\_F606 & S0 & 1.46 & 0.02 & 6.61 & 9 & 65.94 & - & -18.687 & - \\ 
NGC5838  & 28.5 & WFPC2\_F814 & S0 & 0.85 & 0.02 & 8.09 & 12 & 273.56 & - & -20.043 & - \\ 
NGC5866 & 15.35 & ACS\_F814 & S0 & 0.35 & 0.01 & 7.7 & 3 & 162.11 & 0.195 & -20.127 & - \\ 
NGC7457  & 13.24 & WFPC2\_F814 & S0 & 1.13 & 0.03 & 6.95 & 1 & 67.97 & 0.154 & -19.14 & 9.56 \\ 
NGC7743 & 20.7 & WFPC2\_F606 & S0 & 1.49 & 0.02 & 6.59 & 17 & 83.5 & 0.164 & -19.278 & - \\ 
\hline
\end{tabular}
\justify{\text{References. 1: \citet{2019MNRAS.490..600D}, 2: \citet{2010ApJ...717..640P}, 3:  \citet{2006ApJ...647..140F}, 4: \citet{2011AJ....142..167D}, 5: \citet{2016ApJ...830..117M},}
{6: \citet{2013ApJ...764..151G}, 7: \citet{1999AJ....117..744V}, 8: \citet{2019A&A...625A..62T},  9: \citet{2016ApJ...831..134V}, 10: \citet{2016ApJS..222...10S},} 
{11: \citet{2013ARA&A..51..511K}, 12: \citet{2011arXiv1107.2244C}, 13:   \citet{2010ApJ...714..713S}, 14: \citet{2009ApJ...703.1502F}, 15: \citet{2017ApJ...850...15P}, 16: \citet{2019ApJ...872..104N}, 17: \citet{2002ApJ...579..530W}, 18: \citet{2015MNRAS.450.2317S}.}}

\label{tab:table1}
\end{table*}

\begin{table}
\caption{The table lists the best-fitting parameters for the relation \mbox{ x = $\alpha$ CIR + $\beta$ }and correlation coefficients for various relations. N denotes the number of galaxies.}
\begin{adjustbox}{width=\columnwidth}
\label{tab:table2}
\begin{tabular}{cccccc}
\hline
x         & $\alpha$           & $\beta$           & r     & p     & N  \\ \hline
log M$_{BH}$ & -1.33   $\pm$ 0.28    & \ 12.35   $\pm$ 0.27   & -0.74 & > 99.99 & 45 \\
log $\sigma$     & -0.55 $\pm$ 0.08 & \  \ \ 2.78    $\pm$ 0.09 & -0.72 & > 99.99 & 49 \\
M$_{B}$        & \ 2.61  $\pm$ 0.44     & -22.75 $\pm$ 0.45  & \  0.70   & > 99.99 & 46 \\
M$^{*}_{B}$    & -1.46 $\pm$ 0.23    & \ 12.46   $\pm$ 0.22   & -0.84 & \ \ \ 99.99 & 19 \\
\hline
\end{tabular}
\end{adjustbox}
\end{table}


\section{Correlations}
\label{sec:correlations}
In recent years, much effort has been devoted to estimate and interpret correlations between galaxy parameters as they are recognized as an important tool in the investigation of  formation and evolution of galaxies. \citet{2018MNRAS.477.2399A} noticed that the CIR of ellipticals and classical bulges correlates well with the mass of the central SMBH whereas that of spirals and lenticulars with pseudo bulges deviate. Also they reported significant correlations between CIR and various parameters of the SMBH hosting galaxies such as age, stellar mass and central radio emission from low luminosity AGNs. \citet{2020RAA....20...15A} reported remarkable correlations between properties of nuclear rings and CIR of their host galaxies using a sample of 13 early-type spiral galaxies, suggested use of CIR as a tool to explore co-evolution between various structural and dynamical properties of galaxies. Table \ref{tab:table1} shows properties of sample galaxies along with CIR and associated uncertainties. As can be seen from the Table \ref{tab:table1}, many properties are compiled from different authors,  following different methodologies.  Estimations of central black hole mass of the sample galaxies involved stellar and/or gas dynamics  \citep{2019MNRAS.490..600D,2010ApJ...717..640P,2016ApJ...830..117M,2013ApJ...764..151G,2019A&A...625A..62T,2016ApJ...832L..11M,2016ApJS..222...10S,2010ApJ...714..713S,2009ApJ...703.1502F,
2017ApJ...850...15P,2019ApJ...872..104N,2015MNRAS.450.2317S},  astrophysical masers \citep{2013ARA&A..51..511K},  V band mass to light ratio \citep{1999AJ....117..744V} and spheroidal stellar component methods \citep{2011arXiv1107.2244C}.  In addition, BH mass for six galaxies were calculated from stellar velocity dispersion measurement \citep{2006ApJ...647..140F,2005ApJ...620..113D,2002ApJ...579..530W}. Central velocity dispersion, central Mg$_{2}$ index and absolute B band magnitude (M$_{B}$) were taken from HyperLEDA (\url{http://leda.univ-lyon1.fr}; \citet{2014A&A...570A..13M}). Stellar bulge mass (M$^{*}_{B}$) of the sample galaxies is adopted from \citet{2013ApJ...764..184M} and \citet{2013ARA&A..51..511K}. We believe, the inhomogeneities in the data might have weakened  the correlations involving CIR and hence the scatter observed in these relations may be taken as a conservative upperlimit.  
\subsection{Correlation between CIR and M$_{BH}$}
CIR shows good correlation with the mass of the supermassive black hole residing at the centre of the galaxy (linear correlation coefficient $r$ = -0.74 with a significance, $S$ greater than 99.99$\%$). The observed anti-correlation between the CIR and M$_{BH}$ is presented in Figure \ref{fig:figure1}(a) along with the uncertainties involved in their respective estimations.  Out of 50 galaxies with SMBH estimations in the sample, 45 galaxies are obeying the CIR-M$_{BH}$ relation reported in the paper.  The outliers in this correlation are NGC 0404 (marked 1 in Figure \ref{fig:figure1}(a)), NGC 3377 (2), NGC 4150 (3), NGC 5128 (4) and NGC 5866 (5).  NGC 4150 had a merger with a less massive, gas-rich galaxy that would have triggered recent star formation \citep{2011ApJ...727..115C}.  \citet{2017ApJ...836..237N} reported NGC 0404 as an active galaxy with recent star formation in the nucleus.  NGC 5128, also known as Centaurus A is reported as a rare case of a giant radio galaxy with a large amount of molecular gas recently accreted and with ongoing star formation comparable to that of a spiral galaxy \citep{2019ApJ...887...88E}. The gaseous component in this closest AGN host galaxy is believed to have been replenished recently (a few 10$^{8}$ yr) by gas from an external source, probably the accretion of an HI - rich galaxy \citep{2019ApJ...887...88E}. The nearly complete edge-on inclination of the galaxy NGC 5866 \citep{2009ApJ...706..693L} might have affected the measurement of central intensity ratio.  The strong correlation between CIR and M$_{BH}$ reported by \citet{2018MNRAS.477.2399A} for spheroidal systems is seen to extend in field ETGs also.
\subsection{Correlation between CIR and central velocity dispersion}
The central intensity ratio shows an anti correlation ($r$ = -0.72, $S > 99.99 \%$) with the central velocity dispersion of the host galaxy and is given in Figure \ref{fig:figure1}(b). Central velocity dispersion is a measure of galaxy mass, validated by various studies highlighting the correlation between them \citep[e.g.][]{2016ApJ...832..203Z}. Though the correlation is not surprising, it can be used for a quick estimation of central velocity dispersion of more distant galaxies.  In this case also NGC 5866 (marked 5 in Figure \ref{fig:figure1}(b)) and NGC 5128 (marked 4 in Figure \ref{fig:figure1}(b)) are showing deviation from the main trend.
In the case of the third outlier IC 3773 (marked 6 in Figure \ref{fig:figure1}(b)), the central spheroidal component is known to behave not like a bulge \citep{2012ApJS..198....2K}.
Overwhelming evidence has shown that mass of super massive black hole at the centre of the galaxy is well correlated with the velocity dispersion of the host's bulge \citep[e.g.][]{2013ARA&A..51..511K,2016ApJ...818...47S,2019MNRAS.487.5764T,2019GReGr..51...65Z}. This strongly suggests a causal connection between the formation and evolution of the black hole and the bulge. These correlations suggest that CIR of galaxies hold high potential  in understanding the evolution of galaxies. 
\subsection{Correlation between CIR and M$_{B}$}
We find that CIR of the sample galaxies shows an anti-correlation ($r$ = 0.70, $S > 99.99 \%$ ) with the absolute B band magnitude of the host galaxy as shown in Figure \ref{fig:figure1}(c). In this case, six galaxies ( IC 3773 (marked 6 in Figure \ref{fig:figure1}(c)) NGC 0404 (1), NGC 0855 (7), NGC 1316 (8),  NGC 2787 (9) and NGC 5866 (5)) are showing deviation from the linear relation. Edge-on like morphology of NGC 5866, IC 3773 and NGC 0855 may affect the estimation of their CIR.
Deep photometric observations of the galaxy NGC 1316, another outlier indicate that, it may be in a later phase of mass assembly, where smaller satellites recursively accrete into the galaxy \citep{2017ApJ...839...21I}.  NGC 2787 is reported to have a strong bar aligned in PA = 160$^{\circ}$ and star formation is absent over the whole galaxy extension \citep{2019ApJS..244....6S}.  The galaxy could be experiencing a non-AGN-related quenching process as it is reported to contain a relatively low-mass BH (10$^{7.6}$ M$_{\odot}$) at its centre \citep{2016ApJ...832L..11M}. 
Given the positive correlation between M$_{BH}$ and M$_B$ for elliptical galaxies \citep{2007MNRAS.379..711G}, this relation is not surprising and support the idea of co-evolution of black hole and host galaxy. The correlation between M$_{BH}$ and M$_B$ for our sample is good with \textit{r = -0.75} and significance \textit{S > 99.99$\%$}. 
\subsection{Correlation between CIR and M$^{*}_{B}$}
As  already mentioned in Section \ref{sec:correlations}, stellar bulge mass, M$^{*}_{B}$  for our sample galaxies were taken from \cite{2013ARA&A..51..511K} and \cite{2013ApJ...764..184M}.
\cite{2013ARA&A..51..511K} determined M/L$_{K}$ by averaging mass-to-light ratios obtained from stellar population and stellar dynamics separately and used absolute K band magnitude from 2MASS photometric decomposition to calculate stellar bulge mass.
\cite{2013ApJ...764..184M} calculated spherical bulge mass by multiplying V-band luminosity with mass-to-light ratio obtained using spherical Jeans models to fit stellar kinematics \citep{2004ApJ...604L..89H}. Figure \ref{fig:figure1}(d) shows the variation of CIR with M$^{*}_{B}$. This correlation seems to be very strong with the linear correlation coefficient r = -0.84, with significance $S = 99.99 \%$.
Three galaxies (NGC 2787 (marked 9 in Figure \ref{fig:figure1}(d)), NGC 5128 (4) and NGC 7457 (10)) are outliers in the correlation.  
As the bulge mass of the sample galaxies increases, the central light contribution (in the inner aperture) becomes lesser.  This could also be due to the role of central SMBH which is reported to be proportional to the dynamical bulge mass \citep[e.g.][]{2003ApJ...589L..21M,2004ApJ...604L..89H,2016MNRAS.461L.102S}.  
\subsection{Correlation between CIR and central Mg$_{2}$ index}
Mg$_2$ index \citep{1976ApJ...204..668F} of magnesium is most popular among narrow band spectro-photometric indices. It is present in the spectra of all elliptical galaxies as a very prominent absorption feature about 5200 A$^{\circ}$
\citep{1992AJ....103.1814B}. CIR shows a negative correlation  with the central Mg$_2$ index of galaxies (Spearmann rank order correlation coefficient \textit{r$_s$} = -0.67, $S > 99.99 \%$) and is shown in Figure \ref{fig:figure2}. Galaxies NGC 5866 (marked 5 in Figure \ref{fig:figure2}) and IC 3773 (6) are outliers here also.  Mg is one of the $\alpha$ elements which can be produced in core-collapse supernovae (SNe) with very short life times \citep{2016ApJ...832L..11M}. It is proposed that, a series of supernovae explosions near the centre can produce high Mg abundance \citep{2016MNRAS.463.4396M}.  Galactic winds from the supernovae may suppress the star formation and result in the observed reduction in CIR values. As the strength of element Mg in a galaxy is positively correlated with mass of the central SMBH  \citep{2008MNRAS.390..814K}, cumulative effects of feedback from SMBH and supernovae can be held responsible for the correlation.  
\section{DISCUSSION AND CONCLUSION}
\label{sec:discussion}
In this paper, we determine the central intensity ratio of 52 nearby early-type galaxies and demonstrate the usefulness of CIR in the analysis of formation and evolution of galaxies, manifested through multiple two parameter correlations. CIR shows good correlations with mass of SMBH, central velocity dispersion, absolute B band magnitude M$_B$, stellar bulge mass M$^{*}_{B}$ and central Mg$_2$ index of host galaxy. The simple photometric definition of CIR makes it sensitive to any addition (or subtraction) to the light near the centre of a galaxy. Addition of light is possible in systems with high star formation near the centre while quenching via feedback from the central black hole can reduce the CIR value.

The mass of SMBH enjoys several correlations with many host galaxy properties,  including M$_{BH}$ - $\sigma$  \citep[e.g.][]{2013ARA&A..51..511K,2016ApJ...818...47S,2019MNRAS.487.5764T,2019GReGr..51...65Z},   M$_{BH}$ - M$_B$ \citep{2007MNRAS.379..711G},   M$_{BH}$ - M$^{*}_{B}$ \citep[e.g.][]{2003ApJ...589L..21M,2004ApJ...604L..89H,2016MNRAS.461L.102S} and \mbox{M$_{BH}$ -  Mg$_b$} index  \citep{2008MNRAS.390..814K}.   These correlations suggest a strong connection between the growth of SMBH and the evolution of galaxies.

There is growing evidence that,  energetic outflows caused by gas falling onto SMBH can self regulate the growth of the black hole and star formation rate near the centre by heating the surrounding gas  \citep{2006Natur.442..888S,2012Sci...337..544V}. As the mass of SMBH increases, CIR is found to reduce. Low values of CIR,  as the central black hole grows,  can be taken as an evidence for the proposed self regulating feedback mechanism. Both semi analytic models and cosmological hydro dynamic simulations of galaxy formation require an injection of additional energy to prevent cooling of gas and subsequent formation of stars in galaxies. The relation between the temperature of the hot gas in massive ETGs and mass of central SMBH \citep{2019MNRAS.488L..80M} supports the idea that the black hole feedback is a fundamental mechanism in regulating the formation and evolution of massive galaxies. The close relationship between host galaxy properties and the central black holes, described as co-evolution, can probably begin with the birth of both in the universe \citep{2012arXiv1206.2661S}.

The trend between CIR and central Mg$_{2}$ index of host galaxy exhibited by the sample is new. Analysis of the stellar content of massive ETGs clearly indicates that most of their stars formed early on and little new stars were added in the last 5-8 Gyrs \citep{2020MNRAS.491.3562Z}, often invoking feedback from supernovae \citep{2009MNRAS.394.1131K}.  Galaxies with high Mg abundance in their centres might have experienced a series of supernovae explosions. Though these supernovae had small impact on removing cold gas from the interior of galaxies, the rank-order correlation between CIR and central Mg$_2$ index possibly suggests that, feedback from supernovae might also play a role in suppressing the star formation near the centre of galaxies, even though it is thought to be less efficient in massive galaxies with deeper potential wells \citep{2006Natur.442..888S}.

The correlations shown by the central velocity dispersion, absolute B band magnitude M$_B$ and stellar bulge mass M$^{*}_{B}$ with CIR reflect the possible connection between the mass of the galaxy and its central light distribution. Massive galaxies show a suppression of star formation at their centres. For galaxies with high central velocity dispersion ($\sigma > $ 240 km s$^{-1}$), feedback in most central black holes is strong enough to halt star formation near the centre \citep{2006Natur.442..888S}.

The scatter in all these correlations can be treated as an upper limit as all parameters involved in the correlations with CIR are collected from different studies and included different techniques for their determination.  A homogeneous data of CIR in a single filter for a particular morphology (ellipticals/ lenticulars) could have reduced the scatter even further. The various correlations shown by CIR in field ETGs make it suitable to explore the co-evolution scenario in galaxies. The simple photometric tool has the potential to analyse evolution of galaxies in cluster environment also, provided enough spatial resolution is available. The various two parameter correlations with spectroscopic parameters can be used to get a quick estimation of these parameters. Higher dimensional relations (like the Fundamental plane and photometric plane \citep{2000ApJ...531L.103K, 2000ApJ...533..162K}) can be explored and used in the study of simulation of evolution of galaxies.

\section*{Acknowledgements}
We thank the anonymous reviewer for his/her valuable comments which greatly improved the contents of this paper. KS would like to acknowledge the financial support from INSPIRE program conducted by Department of Science and Technology, Government of India. We acknowledge the use of public data based on observations made with the NASA/ESA Hubble Space Telescope, obtained from the data archive at the Space Telescope Science Institute.  We acknowledge the use of the HyperLeda database  (\url{http://leda.univ-lyon1.fr}) and the NED, NASA/IPAC Extragalactic Database (\url{http://ned.ipac.caltech.edu}), which is operated by the Jet Propulsion Laboratory, California Institute of Technology, under contract with the National Aeronautics and Space Administration. 
\section*{Data availability}
The data underlying this article are collected by the NASA/ESA Hubble Space Telescope which are publicly available from the Mikulski Archive for Space Telescopes (MAST) at the Space Telescope Science Institute (STScI). STScI is operated by the Association of Universities for Research in Astronomy, Inc. under NASA contract NAS 5-26555.


\bibliographystyle{mnras}
\bibliography{mybib} 








\bsp	
\label{lastpage}
\end{document}